\documentclass[%
 reprint,
 amsmath,amssymb,
 aps,
]{revtex4-1}

\usepackage{graphicx}

\usepackage{graphicx}
\usepackage{dcolumn}
\usepackage{bm}
\usepackage{amsfonts}
\usepackage{amsmath}

\usepackage{float}
\makeatletter

\newcommand{\Rmnum}[1]{\expandafter\@slowromancap\romannumeral #1@}
\makeatother
\begin{document}

\title{Noncommutative spaces and covariant formulation of statistical mechanics}

\author {V. Hosseinzadeh}\email{v.hosseinzadeh@stu.umz.ac.ir}
\author{M. A. Gorji}\email{m.gorji@stu.umz.ac.ir} \author{K.
Nozari}\email{knozari@umz.ac.ir} \affiliation{Department of Physics,
Faculty of Basic Sciences, University of Mazandaran, P.O. Box
47416-95447 Babolsar, Iran}\author{B.
Vakili}\email{b.vakili@iauctb.ac.ir} \affiliation{Department of
Physics, Central Tehran Branch, Islamic Azad University, Tehran,
Iran}
\begin{abstract}
We study the statistical mechanics of a general Hamiltonian system
in the context of symplectic structure of the corresponding phase
space. This covariant formalism reveals some interesting
correspondences between properties of the phase space and the
associated statistical physics. While topology, as a global
property, turns out to be related to the total number of
microstates, the invariant measure which assigns {\it a priori}
probability distribution over the microstates, is determined by the
local form of the symplectic structure. As an example of a model for
which the phase space has a nontrivial topology, we apply our
formulation on the Snyder noncommutative space-time with de Sitter
four-momentum space and analyze the results. Finally, in the
framework of such a setup, we examine our formalism by studying the
thermodynamical properties of a harmonic oscillator system.

\begin{description}
\item[PACS numbers] 05.20.-y; 02.40.-k; 02.40.Gh
\item[Key Words]
Statistical mechanics, Noncommutative geometry, Symplectic geometry
\end{description}
\end{abstract}
\maketitle
\section{Introduction}
Statistical mechanics is a bridge between mechanics and
thermodynamics. Indeed, it may be assumed as the microscopic
mechanical basis for the macroscopic thermodynamical properties of a
physical system. The key concept which links these two arenas is the
notion of the phase space, that is the space of all distinct
possible microstates. The dynamical evolution is nothing but a
smooth transition from one microstate to another and statistical
mechanics is about how to count the number of these microstates. In
a more technical language, the phase space of a dynamical system is
a symplectic manifold which naturally is equipped with a symplectic
structure \cite{symplectic}. The symplectic structure covariantly
determines the Poisson brackets and Liouville volume and the
kinematics is then defined in a coordinate-independent manner on the
phase space. The dynamics will be also specified upon taking the
Hamiltonian function as the generator of time evolution in this
setup. Therefore, the Hamiltonian determines how the system evolves
through the microstates and the Liouville volume specifies how one
can count and measure them. Consider, for instance, a mechanical
system consisting of $N$ particles which are subjected to some
forces. Each particle moves in Euclidean three-dimensional space
${\mathbf R} ^3$ and the corresponding configuration space is then
${\mathbf R}^{3N}$. Therefore, the phase space will be the cotangent
bundle of this space, that is, ${\mathbf R}^{3N}\times{\mathbf
R}^{3N}$, which naturally admits a symplectic structure. In terms of
the positions and momenta of the particles, the associated
symplectic structure takes the standard canonical form. Such a
canonical symplectic structure, which is also dynamically invariant,
may be used to determine the fundamental canonical Poisson brackets
and the kinematics for the system under consideration. Subsequently,
based on the canonical symplectic structure one leads to the
canonical measure which can be viewed as a mathematical expression
for the principle of equal probabilities in standard statistical
mechanics \cite{laplace}. Although it might be thought that the
symplectic formulation is nothing but a covariant reformulation of
the standard statistical mechanics, its advantages may become more
clear when one is dealing with the Hamiltonian systems with
nontrivial structures (for instance when the standard ${\mathbf R}
^n$ topology of the phase space is replaced with a more complicated
topology or the standard canonical Poisson algebra is replaced with
a deformed noncanonical one). Such systems, are investigated in the
context of doubly special relativity theories, as a flat limit of
quantum gravity, where the four-momentum space has a curved (de
Sitter or anti-de Sitter) geometry with nontrivial topology
\cite{Curve-M}. The space-time structure then turns out to be
noncommutative and the associated Poisson algebra between the
positions and momenta become noncanonical \cite{Girelli,DOS-Ads}.
The space-time with noncommutative coordinates was first proposed by
Snyder in Ref. \cite{Snyder} in order to regularize quantum field
theory in the ultraviolet regime \cite{Snyder-UV}. In more recent
times, it was also used by the stability theory of the Lie algebras
to be a minimal candidate for the relativistic quantum mechanics
\cite{Mendes} and by string theory at a fundamental level
\cite{String-NC}. The duality between curved momentum space and
noncommutative structure was first pointed out by Majid in Ref.
\cite{Majid}. The direct consequence of space-time with
noncommutative coordinates is the modification to the dispersion
relation and the associated density of states
\cite{Mendes-DOS,Girelli2,Glikman,R-L,DOS,DSR-THR}. In symplectic
formulation, the density of states is properly determined by the
Liouville measure which will be constructed from the symplectic
structure \cite{Gorji}. In noncommutative space-time, however, the
corresponding symplectic structure seems to be noncanonical which
induces a nonuniform probability distribution over the set of
microstates through a deformed Liouville volume element. This
dynamically invariant Liouville volume enters in the definition of
the Gibbs entropy and partition function from which all the
thermodynamical properties of a system can be extracted. Therefore,
the topology of the phase space will play an important roll since
all integrals are evaluated over the entire phase space. For
instance, let us consider a system which has an ultraviolet cutoff
due to the existence of a maximal momentum. It is easy to see that
the topology of the spatial sector of the momentum space may be
compact. Moreover, systems with a totally compact phase space, which
can be considered as the classical limit of quantum systems with
finite dimensional Hilbert space (e.g, angular momentum for systems
with fixed total angular momentum) were recently studied in the
context of loop quantum gravity \cite{Rovelli}. For such a compact
phase space, the Liouville volume is finite and consequently a
finite number of microstates naturally arises \cite{LT}. This shows
the correspondence between the topology of the phase space and the
number of microstates which can be realized when the symplectic
geometry is implemented. The black hole physics is a good example in
which the system obeys the thermodynamical laws and its statistical
origin is not thoroughly formulated yet \cite{BH-ST,BH-LQG} in the
framework of Hamiltonian formalism (see also Ref.
\cite{BH-Hamiltonian}).

In this regard, formulating the statistical mechanics in its most
fundamental form is important, at least, for two reasons: (i)
finding a more precise and fundamental interpretation for the basis
of the statistical mechanics, and (ii) studying the statistical
mechanics for the Hamiltonian systems with nontrivial structure,
such as those we have mentioned above. Motivated by the stated
issues, in this paper we are going to formulate statistical
mechanics for a general Hamiltonian system in a covariant
(coordinate-independent) manner. The paper is organized as follows.
In section \Rmnum{2}, we deal with the kinematics and dynamics of a
many-particle system with general Hamiltonian structure in the
context of symplectic geometry. The statistical mechanics for such
systems is formulated in section \Rmnum{3}. In section \Rmnum{4}, by
means of the constructed setup we study the statistical mechanics in
the Snyder noncommutative space-time with curved energy-momentum
space as an explicit example of a physical system with nontrivial
(noncommutative) Hamiltonian structure. As a case study, the
thermodynamical properties of a harmonic oscillator system in Snyder
space is also presented at the end of this section. Section
\Rmnum{5} is devoted to the summary and conclusions.
\section{Hamiltonian systems}
In this section, in order to study the statistical mechanics in the
context of symplectic geometry, we consider the kinematics and
dynamics of an $N$-particle system.
\subsection{Kinematics}
Let us consider a system of $N$ particles similar to conventional
systems in statistical mechanics. The corresponding $6N$-dimensional
kinematical phase space may be obtained by coupling $N$
single-particle phase spaces $\Gamma_\alpha$ with $\alpha=1,...,N$
as
\begin{eqnarray}\label{gamma split n}
\Gamma=\Gamma_1\times...\times\Gamma_N\,.
\end{eqnarray}
The phase space $\Gamma$ naturally admits a symplectic structure
$\omega$ which is a nondegenerate closed 2-form \cite{symplectic}.
The classical state of this $N$-particle system is determined by a
point in $\Gamma$ (phase point). The evolution of phase points is
then determined by the Hamiltonian vector field ${\mathbf x}_{_H}$,
whose integral curves are trajectories of the phase points (phase
trajectories). From the fact that $\omega$ is nondegenerate, one can
assign a vector field to a function $f$ as $\omega({\mathbf
x}_{_f})=df$ and the Poisson bracket between two observables
(real-valued functions on the phase space) can be defined as
\begin{equation}\label{PBD1}
\{f,\,g\}=\omega({\mathbf x}_{_f},{\mathbf x}_{_g})\,.
\end{equation}
These functions with the above structure constitute a Lie algebra
under Poisson brackets on $\Gamma$. We also have the so-called
Liouville volume on $\Gamma$ with its standard definition as
\begin{eqnarray}\label{Vol-D}
\omega^{3N}=\frac{1}{(3N)!}\,\omega\wedge...\wedge\omega
\hspace{0.25cm}(3N\,\,\mbox{times})\,.
\end{eqnarray}
In terms of positions and momenta $x=({\bf q},{\bf p})$ of the
particles, the Liouville volume takes the local form
\begin{align}\label{Vol-D2}
\omega^{3N}=\sqrt{\det\boldsymbol{\omega}}\,dq^1_1\wedge...\wedge
dq^3_N\wedge dp^1_1\wedge...\wedge dp^N_3\equiv \nonumber \\
\sqrt{\det \boldsymbol{\omega}}\,d^{3N}q\,d^{3N}p\,,
\end{align}
where $q^i_\alpha$ and $p_i^\alpha$ are the $i$'th component of the
positions and momenta of the $\alpha$th particle, respectively, and
$\boldsymbol{\omega}$ is the matrix representation of the symplectic
structure $\omega$ with components $\boldsymbol{\omega}_{ij}=\omega(
{\partial}/{\partial{x}^i_\alpha},{\partial}/{\partial x^j_\beta})$.

Now, one may be able to decompose the symplectic structure as
\begin{eqnarray}\label{decomposition}
\omega=\sum_{\alpha=1}^N {\omega}_{\alpha}\,,
\end{eqnarray}
where ${\omega}_{\alpha}$ is the symplectic structure on
$\Gamma_{\alpha}$. Although this assumption has no mathematical
basis, from a physical point of view it is acceptable since the
particles are assumed to be kinematically separated. In other words,
the Lie algebras of the functions defined by $\{\omega_\alpha\}$
over $\{\Gamma_\alpha\}$ are separately closed. Therefore, the
Liouville volume over $\Gamma$ takes the form
\begin{align}\label{decomposition3}
\omega^{3N}=\frac{1}{(3N)!}\,\bigg(\sum_{\alpha=1}^N{\omega}_{\alpha}\bigg)
{\wedge\,.\,.\,.\,\wedge}\bigg(\sum_{\alpha=1}^N{\omega}_{\alpha}\bigg)\,,
\end{align}
where the wedge products take place $3N$ times. The volume element
then works out to be
\begin{align}\label{decomposition 4}
\omega^{3N}=\omega_1^3{\wedge\,...\,\wedge}\omega_N^3=
\bigg(\prod_{\alpha=1}^N\,\sqrt{\det\boldsymbol{\omega}_{\alpha}}\bigg)\,
d^{3N}q\,d^{3N}p\,,
\end{align}
where $\omega^3_{\alpha}$ is the volume element of the corresponding
$\{\Gamma_{\alpha}\}$, and we have also used the fact that
$\omega^n_{\alpha}=0$, for $n>3$. Clearly, $\sqrt{
\det\boldsymbol{\omega}}$, that is, a function on $\Gamma$,
factorizes into products of
$\{\sqrt{\det\boldsymbol{\omega}_{\alpha}}\}$. This cannot be
realized in the most general situation, where the symplectic
structure cannot be written as (\ref{decomposition}).

As a special case, consider a system consisting of $N$ particles
subject to some forces moving in Euclidean three-dimensional space
with standard ${\mathbf R}^3$ topology. Therefore, the phase space
will be ${\mathbf R}^{3N}\times{\mathbf R}^{3N}$ which is indeed the
cotangent bundle of the configuration space, that is, ${\mathbf
R}^{3N}$. Geometrically, the cotangent bundle is a symplectic
manifold endowed with a canonical symplectic structure
\begin{equation}\label{canonical structure}
\omega_c=d{\mathrm{Q}}^i_{\alpha}\wedge d{\mathrm{P}}^{\alpha}_i\,,
\end{equation}
in which $\mathrm{Q}$ and ${\mathrm{P}}$ are interpreted as the
positions and momenta of particles, respectively. The summation is
over both $i=1,2,3$ and $\alpha=1, ...,N$. This is the canonical
representation of symplectic structure and coordinates
$({\mathrm{Q}}^i_{\alpha},{\mathrm{P}}^{\alpha}_i)$ are known as the
canonical coordinates. Substituting the canonical structure
(\ref{canonical structure}) into the definition (\ref{PBD1}), the
Poisson bracket will be
\begin{equation}\label{PBD2}
\{f,\,g\}_{_c}=\frac{\partial f}{\partial {\mathrm{Q}}^i_{\alpha}}
\frac{\partial g}{\partial {\mathrm{P}}^{\alpha}_i}- \frac{\partial
f}{\partial {\mathrm{P}}^{\alpha}_i}\frac{\partial g}{\partial
{\mathrm{Q}}^i_{\alpha}}\,,
\end{equation}
which leads to the standard canonical Poisson algebra
\begin{align}\label{CPA}
\{{\mathrm{Q}}^i_{\alpha},{{\mathrm{Q}}}^j_{\beta}\}_{_c}=0,\hspace{2mm}\{
{\mathrm{Q}}^i_{\alpha},{\mathrm{P}}^{\beta}_j\}_{_c}=\delta^i_j\,\delta^{
\beta}_{\alpha},\hspace{2mm}\{{\mathrm{P}}^{\alpha}_i,{{\mathrm{P}}}^{
\beta}_j\}_{_c}=0.
\end{align}
Also, with the help of (\ref{canonical structure}) we get from
(\ref{Vol-D}) the Liouville volume as
\begin{eqnarray}\label{canonical measure}
\omega_c^{3N}=d^{3N}\mathrm{Q}\,d^{3N}{\mathrm{P}}\,,
\end{eqnarray}
which is nothing but the standard volume element on ${\mathbf
R}^{6N}$. Thus, the kinematics of the standard statistical mechanics
can be recovered as a special case of this setup. Taking into
account the fact that $\det \boldsymbol{\omega_c}=1$, the measure
associated to the standard volume element (\ref{canonical measure})
assigns a uniform probability distribution to the set of
microstates. This is justifiable in the light of Laplace's principle
of indifference, which states that in the absence of any further
information, all outcomes are equally likely \cite{laplace}. This is
the fundamental basis of the statistical mechanics.

In the case where the space-time has a noncommutative structure,
apart from the details of the different models, there is always a
deformation parameter which can be linked to a minimal length
associated with the system under consideration. The direct
consequence of this setup is the modification of the dispersion
relation and the density of states (see Ref. \cite{Mendes} for the
special case of the stable noncommutative algebra for the
relativistic statistical mechanics). The density of states is
determined by the symplectic structure and the associated Liouville
volume. Since the symplectic structure takes a noncanonical form in
terms of the physical positions and momenta $(q,p)$ in
noncommutative spaces \cite{LT}, we have
$\det\boldsymbol{\omega}=\det\boldsymbol{ \omega}(q,p)\neq{1}$. This
result immediately shows that Laplace's indifference principle is no
longer applicable for a noncommutative space-time. This is because
of the fact that there is extra information in these setups (minimal
length or maximal momentum) and the Liouville measure then assigns a
nonuniform probability distribution over the set of microstates by
means of the relation (\ref{Vol-D2}). This simple consideration of
noncommutative space-time as an example of a kinematically deformed
system shows the advantage of the coordinate-independent symplectic
formulation of the statistical mechanics (see Refs.
\cite{Gorji,Photon-Band} for explicit examples). We will explicitly
consider such an example in section IV.

It also should be noted that, according to the Darboux theorem,
there is always a local chart on $\Gamma$ in which any symplectic
structure takes the canonical form with $\det\boldsymbol{\omega}=1$.
In the case of kinematically deformed systems such as the
noncommutative space-time, however, the new canonical coordinates
cannot be interpreted as the positions and momenta of the particles.
Moreover, in the case of standard statistical mechanics, one can
also find a local chart in which the standard canonical structure
$\omega_c$ takes a noncanonical form with
$\det\boldsymbol{\omega_c}\neq{1}$. But again, these new
noncanonical coordinates cannot be interpreted as positions and
momenta of particles. In all of the above cases, as we will see in
the next section, the partition function and consequently resultant
thermodynamical properties are independent of the chart in which the
system is considered.

\subsection{Dynamics}
As we have mentioned above, the time evolution of the system is
determined by the Hamiltonian vector field ${\mathbf x}_{_H}$. It
can be obtained by solving the dynamical equation
\begin{equation}\label{can-dy-eq}
\omega\,({\mathbf x}_{_H})=dH\,,
\end{equation}
where $H$ is the Hamiltonian function of the system that is a
real-valued function over $\Gamma$ and $\omega\,({\mathbf x}_{_H})$
is an interior product of $\omega$ and ${\mathbf x}_{_H}$.
Additionally, ${\mathbf x}_{_H}$ can be expanded as
\begin{equation}
{\mathbf x}_{_H}=\sum_{\alpha=1}^N{\mathbf x}^{\alpha}_{_H}\,,
\end{equation}
where ${\mathbf x}^{\alpha}_{_H}$ is the projection of ${\mathbf
x}_{_H}$ on $\Gamma_{\alpha}$. Note that, in general, the domain of
components of ${\mathbf x}^{\alpha}_{_H}$ is the whole phase space
$\Gamma$ and consequently the Lie brackets between $\{{\mathbf
x}^{\alpha}_{_H}\}$ will not vanish:
\begin{equation}\label{Lie1}
[{\mathbf x}^{\alpha}_{_H},{\mathbf x}^{\beta}_{_H}]\neq0\,.
\end{equation}
However, as a special case, one can consider a system consisting of
noninteracting particles. Therefore, the Hamiltonian function can be
written as the sum of individual Hamiltonians, that is,
\begin{equation}\label{decomposition h}
H=\sum_{\alpha=1}^NH_{\alpha}\,.
\end{equation}
So, from decomposition (\ref{decomposition}) for the symplectic
structure, the dynamical equation (\ref{can-dy-eq}) becomes
\begin{eqnarray}\label{dy eq decomposition}
\bigg(\sum_{\alpha=1}^N {\omega}_{\alpha}\bigg) \bigg(
\sum_{\beta=1}^N {\mathbf x}^{\beta}_{_H} \bigg)=\sum_{\alpha=1}^N
d{H}_{\alpha}\,.
\end{eqnarray}
Using the fact that ${\omega}_{\alpha}({\mathbf x}^{\beta}_{_H})=0$,
for $\alpha\neq\beta$, one is led to a set of $N$ independent
differential equations
\begin{eqnarray}\label{dy eq decomposition 1}
{\omega}_{1}({\mathbf
x}^{1}_{_H})=d{H}_{1}\,,\,\,.\,.\,.\,\,,\,{\omega}_{N} ({\mathbf
x}^{N}_{_H})=d{H}_{N}\,.
\end{eqnarray}
Now, it is clear that the domain of the components of ${\mathbf
x}^{\alpha}_{_H}$ will be $\Gamma_{\alpha}$ and the Lie brackets
between ${\mathbf x}^{\alpha }_{_H}$ and ${\mathbf x}^{\beta}_{_H}$
vanish
\begin{equation}\label{criterion separability}
[{\mathbf x}^{\alpha}_{_H},{\mathbf x}^{\beta}_{_H}]=0\,.
\end{equation}
This can be seen as a criterion for the {\it kinematical and
dynamical separability} of the particles. In the Appendix, we will
see that this benchmark shows itself as a criterion for the
statistical independence of the macroscopic subsystems which in turn
is a definition of equilibrium.

The other important property of the Hamiltonian systems is the
existence of an invariant measure that enables one to construct the
equilibrium statistical mechanics through the well-known Liouville
theorem as
\begin{equation}\label{Liouvillee}
\frac{d}{dt}\,\omega^{3N}=0\,,
\end{equation}
where ${d}/{dt}={\partial}/{\partial t}+{\mathcal{L}}_{{\mathbf
x}_{_H}}$ is the total time derivative and ${\mathcal{L}}_{{\mathbf
x}_{_H}}$ denotes the Lie derivative with respect to ${\mathbf
x}_{_H}$. The conservation of $\omega^{3N}$ can be traced back to
the fact that ${\mathcal{L}}_{ {\mathbf x}_{_H}}\omega^{3N}=3N
({\mathcal{L}}_{{\mathbf x}_{_H}}\omega)\wedge \omega^{3N-1}$ and
${\mathcal{L}}_{{\mathbf x}_{_H}}\omega=d\omega\,({ {\mathbf
x}_{_H}})+d(\omega\,({{\mathbf x}_{_H}}))=0$, where we have used
$d(\omega\,({{\mathbf x}_{_H}}))=d^2H=0$ and the closure of the
symplectic structure $d\omega=0$. It is obvious that $\omega^{3N}$
is not explicitly time dependent. Note also that the Liouville
theorem arises from the particular dynamics of a Hamiltonian system.

Consider the standard statistical mechanics as a special case, where
the kinematics is given by canonical symplectic structure
(\ref{canonical structure}). The dynamical equation
(\ref{can-dy-eq}) then leads to the familiar form of the Hamilton's
equations in terms of the physical positions and momenta as
\begin{equation}\label{H equation}
\frac{{d\mathrm{Q}}^i_{\alpha}}{dt}=\frac{\partial H}{\partial
\mathrm{P}^{\alpha}_i}
\,,\hspace{1cm}\frac{{d\mathrm{P}}^{\alpha}_i}{dt}=-\frac{\partial
H}{\partial \mathrm{Q}^i_{\alpha}}\,.
\end{equation}
Note that when the symplectic structure has a noncanonical form, the
form of Hamilton's equations would deviate from the relation (\ref{H
equation}).

The triple $(\Gamma,\omega,H)$ constitutes a general Hamiltonian
system and we explore the statistical physics of this general system
in the next section by means of an invariant volume element
$\omega^{3N}$ which is covariantly defined in the relation
(\ref{Vol-D}).

\section{Statistical mechanics}
Statistical mechanics links macroscopic properties of a system to
microscopic laws that are quantum mechanical at the fundamental
level. The basic concept is the von Neumann entropy that is defined
as
\begin{eqnarray}\label{von entropy}
S=-\text{Tr}\,(\widehat{\rho}\ln\widehat{\rho})\,,
\end{eqnarray}
where $\widehat{\rho}$ is the density operator \cite{unit}. Using
the principle of maximum entropy and considering specific relevant
statistical constraints, we can find the density operator which
contains all the statistical information about the system at
equilibrium. For example, for a system in contact with a thermal
bath at temperature $T$ and therefore with a fixed mean energy, the
resultant normalized density operator works out to be
\begin{eqnarray}\label{can rho}
\widehat{\rho}_{_c}=\frac{1}{\mathcal{Z}}\exp[-\widehat{H}/T]\,,
\end{eqnarray}
where $\widehat{H}$ is the Hamiltonian operator of the system and
$\mathcal{Z}$ is called the canonical partition function which is
defined as
\begin{eqnarray}\label{QPF}
{\mathcal
Z}=\text{Tr}\exp[-\widehat{H}/T]=\sum_{i}\exp\big[-\varepsilon_i/T
\big]\,,
\end{eqnarray}
where $\{\varepsilon_i\}$ are the energy eigenvalues of
$\widehat{H}$ and the summation is taken over all the accessible
microstates for the system. At the fundamental level, the
microscopic laws are quantum mechanical and the microstates are
determined by quantum mechanics. On the other hand, in the classical
limit, the microstate of a system represents itself as a point on
the corresponding phase space and due to the infinite resolution of
phase space points, the summation over microstates is really not
well defined. Consequently, one cannot construct a full classical
statistical mechanics in essence. However, quantum mechanics
provides the semiclassical approximation in the spirit of the
uncertainty principle. Indeed, a phase space with finite resolution
and well-defined number of classical microstates can be achieved
through the approximation \cite{Gorji}
\begin{eqnarray}\label{Approx}
\text{Tr}\rightarrow\frac{1}{N!}\int_{\Gamma}\omega^{3N}\,,
\end{eqnarray}
where $1/(N!)$ is due to the fact that the particles may be
indistinguishable and $\omega^{3N}$ is the dynamically invariant
Liouville volume element of the $6N$-dimensional symplectic manifold
$\Gamma$ which is defined in the relation (\ref{Vol-D}). It is
important to note that the semiclassical approximation
(\ref{Approx}) coincides with the full quantum consideration at high
temperature regime. In this limit, the von Neumann entropy (\ref{von
entropy}) is replaced with the Gibbs entropy
\begin{eqnarray}\label{gibbs entropy}
S=-\frac{1}{N!}\int_{\Gamma}\rho \ln{\rho}\,\omega^{3N}\,,
\end{eqnarray}
where $\rho$ is now interpreted as a probability density over the
space of microstates $\Gamma$. The maximum entropy principle for a
system in heat bath at temperature $T$ leads to the Gibbs state
\begin{eqnarray}\label{can rho class}
\rho_c=\frac{1}{\mathcal{Z}}\exp[-H/T]\,,
\end{eqnarray}
where $H$ is the Hamiltonian function of the system and
$\mathcal{Z}$ is the semiclassical canonical partition function
\begin{eqnarray}\label{partition f}
{\mathcal
Z}=\frac{1}{N!}\int_{\Gamma}\exp\big[-H/T\big]\omega^{3N}\,.
\end{eqnarray}
The result of the integral over the phase space for the total
partition function (\ref{partition f}) is independent of a chart in
which the system is considered when the dynamical equation
(\ref{can-dy-eq}) is satisfied in a chart-independent manner. While
the local form of the symplectic structure determines the
probability distribution over the set of microstates, from the
statistical point of view, we expect that the resultant
thermodynamical properties extracted from a partition function are
independent of a chart in which the physical system is considered
(see, for instance, Ref. \cite{Gorji} in which the partition
function in two different charts is calculated).

By considering the decompositions (\ref{decomposition}) and
(\ref{decomposition h}) for the symplectic structure $\omega$ and
Hamiltonian function $H$, respectively, the probability density will
also be decomposed as
\begin{eqnarray}\label{decomposition 2}
\rho_c\,\omega^{3N}=(\rho_c^1\,\omega^3_1)\wedge.\,.\,.\wedge
(\rho_c^N\,\omega^3_N)\,,
\end{eqnarray}
where $\rho_c^\alpha$ is the Gibbs state defined on $\Gamma_\alpha$.
Hence the integral over $\Gamma$ can be factorized into the products
of the integrals over the phase spaces of the particles as
\begin{eqnarray}\label{partition fgh}
{\mathcal Z}=\frac{1}{N!}\prod_{\alpha=1}^{N}\bigg(\int_{\Gamma_{
\alpha}}\exp\big[-H_{\alpha}/T\big]\,\omega^3_{\alpha}\bigg)\,.
\end{eqnarray}
Now, if the particles are influenced by the same kinematics and
dynamics, the partition function can be written in the well-known
form
\begin{eqnarray}\label{partition fghi}
{\mathcal Z}=\frac{{\mathcal Z}_1^N}{N!}\,,
\end{eqnarray}
where we have defined the single-particle state partition function
\begin{eqnarray}\label{partition-sps}
{\mathcal Z}_1=\int_{\Gamma_1}\exp[-H_1/T]\,\omega_1^3\,.
\end{eqnarray}
It is usually claimed that the relation (\ref{partition fghi}) is
applicable when one ignores the quantum correlations between
particles. This means that decompositions such as
(\ref{decomposition}) and (\ref{decomposition h}) may be considered
as conditions for the absence of quantum correlations. Moreover, the
general formalism which is formulated in this section, can also
support a Hamiltonian system with very nontrivial structure, {\it
e.g.} a kinematically classical entangled system in which the
decomposition (\ref{decomposition}) is no longer applicable.

\section{Application in noncommutative Spaces}

Existence of a minimal length is suggested by any quantum gravity
candidate such as string theory and loop quantum gravity
\cite{String,LQG}. It is then widely believed that a
nongravitational theory which supports the existence of a minimal
length would arise at the flat limit of quantum gravity. Such a
theory, would be reduced to the standard relativistic quantum
mechanics at low energy regime where the effects of a minimal length
are negligible. Evidently, taking a minimal length into account
naturally leads to the space-time with noncommutative coordinates.
The first attempt in this direction was done by Snyder in 1947 who
formulated a Lorentz-invariant discrete space-time with
noncommutative coordinates \cite{Snyder}. The space-time with
noncommutative structure is also suggested by the theory of
stability of the Lie algebras \cite{Mendes}. On the other hand,
existence of a universal minimal length cannot be supported by the
special relativity since any length scale in one inertial frame may
be different in another observer's frame through the well-known
Lorentz-Fitzgerald contraction. Thus, the doubly special relativity
theories are formulated in order to take into account an
observer-independent minimum length as well as the velocity of light
\cite{DSR}. The curved four-momentum space then naturally appears to
be a suitable framework to formulate the doubly special relativity
theories \cite{Curve-M}. Furthermore, the different doubly special
relativity theories can be obtained from the different basis of the
$\kappa$-Poincar\'{e} algebra on the associated $\kappa$-Minkowski
noncommutative space-time \cite{Poincare}. In this respect, it is
also shown that the Snyder noncommutative algebra which is proposed
in Ref. \cite{Snyder} and the stable algebra of relativistic quantum
mechanics that is obtained in Ref. \cite{Mendes} can be obtained
from a ten-dimensional phase space with curved geometry for the
four-momentum space through the symplectic reduction process
\cite{Girelli}. Thus, it seems that the space-time with curved
four-momentum space and noncommutative coordinates is a fundamental
framework for taking into account a minimal length scale in the flat
limit of quantum gravity \cite{F-QG}. It would also be claimed that
the Lorentz invariance is an approximate symmetry and will be broken
at high energy regime. Therefore, the Lorentz-violating
noncommutative algebra was proposed by Camelia in Ref. \cite{DSR}
which supports the existence of a minimal observer-independent
length scale. Magueijo and Smolin showed that the Lorentz invariance
can also be preserved by a nonlinear action of the Lorentz group on
the momentum space \cite{DSR-LI}. In order to preserve the Lorentz
invariance, the following commutation relations for the generators
will be held
\begin{align}\label{Algebra}
&\{J_{ab},J_{cd}\}=J_{ad}\eta_{bc}+J_{bc}\eta_{ad}-
J_{bd}\eta_{ac}-J_{ac}\eta_{bd},\\
&\{J_{ab},\,p_c\,\}\,=p_a\eta_{bc}-p_b\eta_{ac},\hspace{1mm}
\{J_{ab},\,x_c\,\}\,=x_a\eta_{bc}-x_b\eta_{ac},\nonumber
\end{align}
where $\eta_{ab}=\mbox{diag}(+1,-1,-1,-1)$ is the Minkowski metric
and $a,b,c,d=0,...,3$. The commutation relations between $(x_a,p_a)$
then classify the different Lorentz-invariant algebras such as the
Snyder noncommutative algebra \cite{Snyder}, the stable algebra
\cite{Mendes} and also the standard Lorentz algebra.

While the relativistic algebras are written in the eight-dimensional
relativistic phase space, to formulate the statistical mechanics one
needs to work in a six-dimensional nonrelativistic subalgebra of the
deformed relativistic algebras in which the time parameter is fixed
and the corresponding Poisson brackets take a noncanonical form. By
means of the resulting deformed Hamiltonian system, one can study
the statistical mechanics with the help of the constructed setup in
the pervious sections. In recent years, many works have gone in this
direction in order to study the effects of an ultraviolet cutoff
(minimal length and maximal momentum) on the thermodynamical
properties of the physical systems. For instance, thermodynamics of
some physical systems in noncommutative spaces is studied in Ref.
\cite{NC-THR}. For the special case of the Snyder noncommutative
space, see Ref. \cite{Snyder-Ther}, in which the thermodynamics of
the early Universe  is explored. It is shown that the Liouville
theorem for the deformed phase space with general noncanonical
Poisson algebra is satisfied. However, as we have shown in section
\Rmnum{3}, the Liouville theorem can be justified in a very simple
manner in the covariant formalism, which also shows the advantage of
the setup. Furthermore, thermodynamical ideal gases are studied in
the context of doubly special relativity theories \cite{DSR-THR}.
Also, motivated by the string theory, the generalized uncertainty
principle is suggested which supports the existence of a minimal
length as a nonzero uncertainty in position measurement
\cite{String,GUP}. For the statistical mechanics in this setup, see
Refs. \cite{GUP-THR,Fityo}. Inspired by loop quantum gravity, the
polymer quantum mechanics is formulated which supports the existence
of a minimal length known as the polymer length scale \cite{QPR}.
Thermodynamical properties of the ideal gases and harmonic
oscillator are also studied in Refs. \cite{Gorji,PL-THR}. To see the
effects of minimal length on the thermodynamics of the black holes
in noncommutative space, the generalized uncertainty principle
framework, and the polymer quantization scheme, see Refs.
\cite{BH-NC,BH-GUP,BH-PL}. Apart from the details of the
above-mentioned effective approaches to quantum gravity
phenomenology, all of them try to consider the effects of a minimal
length in a relevant manner, and implementing the covariant
formalism can clarify the consequences and applications in its most
fundamental form in the language of symplectic geometry.

\subsection{Statistical mechanics in the Snyder noncommutative space}
As we have mentioned above, the advantages of the covariant
formulation may be revealed when a Hamiltonian system with
nontrivial structure is considered. Therefore, in this subsection we
study the statistical mechanics in the Snyder noncommutative space
as a well-known example of a kinematically deformed Hamiltonian
system.

The Snyder relativistic algebra preserves the Lorentz invariance and
therefore the commutation relations (\ref{Algebra}) are satisfied by
the corresponding generators. The commutation relations between
$(x_a,p_a)$ which define the Snyder algebra are given by
\cite{Snyder}
\begin{align}\label{Snyder-Algebra}
\{x_a,x_b\}=\,\frac{J_{ab}}{\kappa^2},\hspace{1mm}\{x_a,p_b\}=
\eta_{ab}+\,\frac{p_ap_b}{\kappa^2},\hspace{1mm}\{p_a,p_b\}=0,
\end{align}
where $\kappa$ is the deformation parameter with the dimension of
the inverse of length which plays the role of the universal quantum
gravity scale in this setup. The four-momentum space of the
eight-dimensional relativistic phase space of a test particle which
moves on the space-time with noncommutative algebra
(\ref{Snyder-Algebra}), is a de Sitter space with topology ${\mathbf
R} \times{\mathbf S}^3$ \cite{Snyder,Girelli}. Identifying the
energy space with ${\mathbf R}$, the topology for the space of
spatial momenta will be ${\mathbf S} ^3$. In order to study the
statistical mechanics in this setup, one should consider the
nonrelativistic subalgebra of (\ref{Snyder-Algebra}) and replace the
standard canonical Poisson algebra with them. The nonrelativistic
Snyder algebra for a particle moves in three-dimensional Euclidean
space ${\mathbf R}^3$ and is then given by \cite{Mignemi}
\begin{align}\label{Snyder-Algebra-NR}
&\{q_i,q_j\}=\,\frac{J_{ij}}{\kappa^2},\hspace{1mm}\{q_i,p_j\}=
\delta_{ij}+\,\frac{p_ip_j}{\kappa^2},\hspace{1mm}\{p_i,p_j\}=0.
\end{align}
It is straightforward to show that the above commutation
relations can be realized from the symplectic structure
\cite{Snyder2}
\begin{eqnarray}\label{Snyder-symplectic}
\omega=dq^i\wedge dp_i-\frac{1}{2}\,d(q^ip_i)\wedge{d}\ln\big(
p^2+\kappa^2\big),
\end{eqnarray}
through the covariant definition (\ref{PBD1}) for the Poisson
brackets. The associated phase space of the particle is then simply
the cotangent bundle of the configuration space which now has the
nontrivial ${\mathbf R}^3\times{\mathbf S}^3$ topology. Indeed, the
space of the spatial momenta has compact ${\mathbf S}^3$ topology
rather than the standard ${\mathbf R}^3$ one (see Refs.
\cite{LT,Mignemi,Mignemi-DOS} for more details). Since the quantum
gravity cutoff (which is defined by the deformation parameter
$\kappa$) is universal, it will be the same for all of the
particles. Thus, for the kinematically deformed system consisting of
$N$ particles which obeys the Snyder noncommutative algebra
(\ref{Snyder-Algebra-NR}), the total symplectic structure can be
obtained by the relation (\ref{decomposition}). The phase space
$\Gamma_\alpha$ for all of the particles is the same and has the
nontrivial ${\mathbf R}^3\times {\mathbf S}^3$ topology. The
associated total phase space $\Gamma_{\rm S}$ can be easily obtained
through the coupling (\ref{gamma split n}), which has the following
nontrivial topology:
\begin{equation}\label{Snyder-gamma}
{\mathbf R}^{3N}\times\,\underbrace{{\mathbf S}^3\times\,...\,
\times{\mathbf S}^3}_{\text{N times}}\,.
\end{equation}
Substituting the symplectic structure
(\ref{Snyder-symplectic}) into the definition (\ref{Vol-D}),
the corresponding Liouville volume takes the form
\begin{align}\label{Snyder-Vol}
\omega^{3N}=\prod_{\alpha=1}^{N}\bigg(\frac{d^3q_\alpha\wedge\,
d^3p^\alpha}{\big(1+(p^\alpha/\kappa)^2\big)}\bigg)=\frac{d^{
3N}q\wedge\,d^{3N}p}{\prod_{\alpha=1}^{N}\big(1+(p^\alpha/
\kappa)^2\big)}\,,
\end{align}
where $q^i_\alpha$ and $p_i^\alpha$ are the $i$'th component of the
positions and momenta of the $\alpha$th particle respectively. From
this relation it is clear that the probability distribution is
nonuniform in Snyder noncommutative space and the standard uniform
one can be recovered only at low energy (low temperature) regime
$\kappa \rightarrow\,\infty$. Indeed, the existence of a minimal
length, as an extra information, changes the probability
distribution at the high energy regime. Thus, in all of the
kinematically deformed noncommutative phase spaces, the local form
of the symplectic structure (in terms of the positions and momenta
of the particles) is noncanonical and therefore the probability
distribution will be nonuniform such that the microstates with
higher momenta are less probable. This is the advantage of the
symplectic covariant formulation of the statistical mechanics which
shows that the probability distribution will be nonuniform on any
phase space with noncanonical Poisson algebra like the Snyder
algebra (\ref{Snyder-Algebra-NR}). However, in light of the Darboux
theorem, one can always find a local chart through the Darboux
transformation $(q^i_\alpha,p_i^\alpha)
\rightarrow({X}^i_\alpha=q^i_\alpha-\frac{q^j_{\alpha}
p_i^\alpha}{(p^\alpha)^2+\kappa^2}\,\delta^{ik}p_k^\alpha,
{Y}_i^\alpha=p_i^\alpha)$, in which the symplectic structure
(\ref{Snyder-symplectic}) takes the canonical form $\omega_{
\alpha}=d{X}^i_\alpha{\wedge}d{Y}_i^\alpha$ for all of the
particles. Thus, the associated Liouville volume becomes
$\omega^{3N}=d^{3N}X\wedge\,d^{3N}Y$, which induces a uniform
probability distribution over the set of microstates. But, it is
important to note that $({X}^i_\alpha,{Y}_i^ \alpha)$ are different
from the standard canonical ones $(Q^i_\alpha,P_i^\alpha)$ which are
defined in relations (\ref{canonical structure}) and (\ref{canonical
measure}). Indeed, while $(Q^i_\alpha,P_i^\alpha)$ are the positions
and momenta of the particles by definition,
$({X}^i_\alpha,{Y}_i^\alpha)$ cannot be interpreted as the positions
and momenta of the particles. Also, the noncanonical Snyder
Liouville volume (\ref{Snyder-Vol}) reduces to the standard
canonical one (\ref{canonical measure}) in the limit of
$\kappa\rightarrow \infty$ and the noncanonical variables
$(q^i_\alpha,p_i^ \alpha)$ then coincide with standard canonical
variables $(Q^i_\alpha,P_i^\alpha)$ in this limit. However, the
canonical variables $({X}^i_\alpha,{Y}_i^\alpha)$ are obtained
through a Darboux transformation and the functional form of the
Hamiltonian function will be modified (compared with the standard
functional form) in this chart as $H(X, Y)$ since the dynamical
equation (\ref{can-dy-eq}) will be satisfied in a chart-independent
manner.

Substituting the Liouville volume (\ref{Snyder-Vol}) in the relation
(\ref{partition f}), the total partition function will be
\begin{eqnarray}\label{Snyder-TPF}
{\mathcal Z}=\frac{1}{N!}\int_{\Gamma_{\rm S}}\frac{\exp\big[
-H(q,p)/T\big]}{\prod_{\alpha=1}^{N}\big(1+(p^\alpha/\kappa)^2
\big)}d^{3N}qd^{3N}p\,.
\end{eqnarray}
Since the Snyder Liouville volume (\ref{Snyder-Vol}) is decomposed
similar to the relation (\ref{decomposition 4}), for the
noninteracting systems in which the total Hamiltonian function also
decomposes as the relation (\ref{decomposition h}), one can rewrite
the total partition function (\ref{Snyder-TPF}) in terms of the
single-particle partition function. Taking the decomposition
(\ref{decomposition h}) into account and substituting the volume
(\ref{Snyder-Vol}) into relation (\ref{Snyder-TPF}), the total
partition function can be rewritten in the form of the relation
(\ref{partition fgh}) as
\begin{eqnarray}\label{Snyder-D-TPF}
{\mathcal
Z}=\frac{1}{N!}\prod_{\alpha=1}^{N}\bigg(\int_{\Gamma_{\alpha}
}\frac{\exp\big[-H_{\alpha}(q_\alpha,p^\alpha)/T\big]}{\big[1+(p^\alpha
/\kappa)^2\big]}\,d^3q_{\alpha}d^3p^\alpha\bigg)\nonumber
\\=\frac{1}{N!} {\mathcal Z_1}^N,\hspace{6cm}
\end{eqnarray}
where we have defined the single-particle partition function as
\begin{equation}\label{Snyder-PF1}
{\mathcal Z_1}=\int_{{\mathbf R}^3}d^3q_1\int_{{\mathbf
S}^3}d^3p^1\frac{
\exp\big[-H_1(q_1,p^1)/T\big]}{\big(1+(p^1/\kappa)^2\big)}\,,
\end{equation}
and we have also assumed that the Hamiltonian functions are the same
for all of the particles. Then, all the thermodynamical properties
of a system can be deduced from the total partition function
(\ref{Snyder-TPF}) or (\ref{Snyder-D-TPF}) in which, as we have
mentioned in the pervious section, the Gibbs factor $1/N!$ is
considered for the nonlocalized systems such as the ideal gas and it
should then be removed for the localized systems such as the
harmonic oscillator which is the subject of the next subsection.
\subsection{Thermodynamics of 3D harmonic oscillator}
Now let us consider a system of $N$ independent and similar
three-dimensional harmonic oscillators. Since the oscillators are
assumed to be independent, the relations (\ref{decomposition}) and
(\ref{decomposition h}) will be satisfied and one then can implement
the relation (\ref{Snyder-D-TPF}) in order to obtain the partition
function of the system. The Hamiltonian of the three-dimensional
simple harmonic oscillator is given by $H_1(q,p)=\frac{p^2}{2m}+
\frac{1}{2}m\sigma^2q^2$, where $m$ is the mass of the oscillator,
$\sigma$ is the frequency, $p^2=p_x^2+p_y^2+p_z^2$ and
$q^2=x^2+y^2+z^2$. Substituting this Hamiltonian into the relation
(\ref{Snyder-PF1}), gives the following single-particle partition
function
\begin{align}\label{HO-pf}
{\mathcal Z_1}=\frac{\kappa^2T^2}{m\hbar^3\sigma^3}\left(1-\frac{
\sqrt{\pi}\kappa}{\sqrt{2mT}}{\mbox{erfc}\bigg[\frac{\kappa}{
\sqrt{2mT}}\bigg]\exp\bigg[\frac{\kappa^2}{2mT}\bigg]}\right)\,.
\end{align}
At the high temperature regime, the second term in the right-hand
side of the above relation is negligible. Under this condition the
dominant contribution is due to the first term and the partition
function is approximated by ${\mathcal
Z_1}\approx\frac{\kappa^2T^2}{m\hbar^3\sigma^3}$. This shows that in
the domain of the high temperature, the number of degrees of freedom
will be reduced since the partition function is proportional to
$\sim{T^2}$ rather than the standard one which is proportional to
$\sim{T^3}$. We will show this feature in a more precise manner in
this subsection. The total partition function can be obtained from
the relation (\ref{Snyder-D-TPF}) as
\begin{eqnarray}\label{HO-TPF}
{\mathcal Z}={\mathcal Z_1}^N,
\end{eqnarray}
where the Gibbs factor $1/N!$ is dropped since the oscillators are
localized. Substituting the single-particle partition function
(\ref{HO-pf}) into this relation, the total partition function for
$N$ three-dimensional harmonic oscillators can be obtained from
which one is able to study the thermodynamical behavior of the
system in Snyder space. However, before doing this task, it is
useful to consider the limit $\frac{\kappa}{ \sqrt{2mT}}\gg1$ which
is related to the first order corrections that would arise from the
minimal length effects. This limit also allows us to compare our
result with the full quantum partition function. In this limit, we
may use the relation
$\mbox{erfc}(x)=\frac{e^{-x^2}}{\sqrt{\pi}x}\sum_{
n=0}^{\infty}(-1)^n\frac{(2n-1)!!}{(2x^2)^n}$, to arrive at the
following asymptotic expression for the partition function
\begin{eqnarray}\label{HO-pf-asym}
{\mathcal Z_1}=\left(\frac{T}{\hbar\sigma}\right)^3\left(1+
\sum_{n=1}^{\infty}(-1)^n\frac{(2n+1)!!}{\left(2\Theta^2
\right)^n}\right)\,,
\end{eqnarray}
in which we have defined the dimensionless variable $\Theta
\equiv\frac{\kappa}{\sqrt{2mT}}$. In order to obtain the quantum
partition function, one should solve the Hamiltonian eigenvalue
problem in Snyder space which is not an easy task at all since the
Schr\"{o}dinger equation takes a complicated form in this setup.
However, in the context of the generalized uncertainty principle the
problem of a $D$-dimensional harmonic is exactly solved \cite{HO-SE}
and the results are applicable also for the Snyder algebra
(\ref{Snyder-Algebra-NR}) as a particular case. The energy
eigenvalues for a three-dimensional harmonic oscillator in Snyder
space are given by
\begin{eqnarray}\label{HO-EE}
E_{nl}=\hbar\sigma\bigg[\Big(n+\frac{3}{2}\Big)\sqrt{1
+\Big(\frac{m\hbar\sigma}{2\kappa^2}\Big)^2}\nonumber
\hspace{3cm}\\+\frac{m\hbar\sigma}{2\kappa^2}\left(
\Big(n+\frac{3}{2}\Big)^2-\Big(l(l+1)+\frac{3}{4}\Big)
\right)\bigg],\hspace{0.5cm}
\end{eqnarray}
where $n=0,1,2,...$, and $0\leq{l}\leq{n}$ and also we have
considered relevant parameter identification to rewrite the result
of the Ref. \cite{HO-SE} in our notation and units. The associated
quantum partition function then will be obtained from the definition
(\ref{QPF}) as
\begin{equation}\label{HO-QPF}
{\mathcal Z}_1=\sum_{n=0}^{\infty}\sum_{l=0}^n\,
\exp\left[-\frac{E_{nl}}{T}\right]\,.
\end{equation}
The above summation cannot be evaluated analytically. However, in
Ref. \cite{Fityo}, it is shown that up to the first order of
approximation, in the limit of $\Theta=\frac{\kappa
}{\sqrt{2mT}}\gg1$, it leads to ${\mathcal Z_1}=
\left(\frac{T}{\hbar\sigma}\right)^3\left(1-\frac{
3}{2}\Theta^{-2}\right)+{\mathcal O}\left[\Theta^{ -4}\right]$,
which is nothing other than the first term in the summation
(\ref{HO-pf-asym}). This coincidence shows that the quantum
partition function (\ref{HO-QPF}) correctly leads to the
semiclassical partition function (\ref{HO-pf}) up to the first order
of approximation at the high temperature regime. It also reveals the
advantages of semiclassical approximation (\ref{Approx}) in the
sense that although there is not an analytical solution for the case
of full quantum partition function (\ref{HO-QPF}), the relation
(\ref{HO-pf}) provides an analytical expression for the high
temperatures. Therefore, in dealing with the statistical
considerations of the physical systems in the presence of a minimal
length, since one is usually interested in the high temperature
regime (such as early Universe thermodynamics), the usage of the
semiclassical approximation seems to be quite reasonable.

\begin{figure}
\flushleft\leftskip+3em{\includegraphics[width=3in]{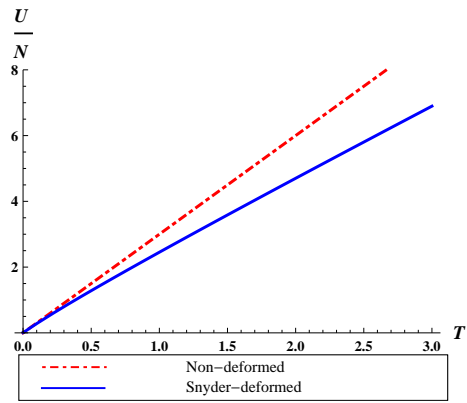}}
\hspace{3cm}\caption{\label{fig:1} Internal energy versus
temperature is plotted. The solid and dashed lines represent the
internal energy for the Snyder-deformed and nondeformed cases,
respectively. The deviation from the standard case arises at the
high temperature regime when the minimal length effects dominate.
The figure is plotted for $m= 1=\kappa$.}
\end{figure}

Substituting the single-particle partition function (\ref{HO-pf})
into the relation (\ref{HO-TPF}) gives the total partition function.
The internal energy then can be obtained from the standard
definition $U=T^2\left(\frac{\partial\ln{\mathcal Z}}{
\partial{T}}\right)_N$ as
\begin{eqnarray}\label{HO-IE}
U=\frac{3NT}{2}\left[1+\frac{1}{3}\left(1-\sqrt{\pi}\,
\Theta\,\mbox{erfc}[\Theta]\,{e}^{\Theta^2}\right)^{
-1}-\Theta^2\right].\hspace{.5cm}
\end{eqnarray}
In figure \ref{fig:1} we have plotted the internal energy versus
temperature in comparison with its nondeformed counterpart. The
specific heat can also be deduced from the definition $C_{_V}=\left(
\frac{\partial{U}}{\partial{T}}\right)_V$ which leads to a somehow
cumbersome expression.

\begin{figure}
\flushleft\leftskip+3em{\includegraphics[width=3in]{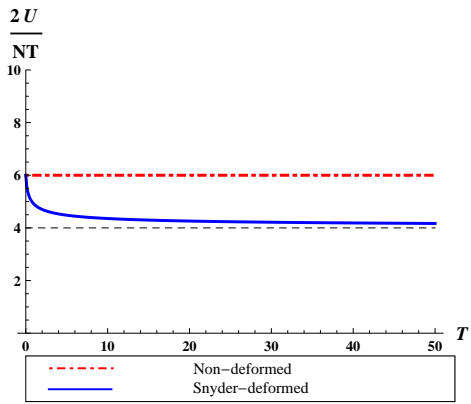}}
\hspace{3cm}\caption{\label{fig:2} The number of degrees of freedom
versus temperature. As it is clear from the figure, in Snyder space
the number of degrees of freedom for a three-dimensional harmonic
oscillator is reduced from 6 to 4 at the high temperature regime.}
\end{figure}

An interesting feature here is the reduction of the number of
degrees of freedom in Snyder space. According to the well-known
equipartition theorem of energy, for the Hamiltonians of the form
$H_1(q,p) =\frac{p^2}{2m}+\frac{1}{2}m\sigma^2q^2$, each of the 6
degrees of freedom makes a contribution of $\frac{1}{2}T$ towards
the internal energy. Thus one may consider $\frac{2U}{NT}$ as the
number of degrees of freedom for each three-dimensional harmonic
oscillator. In figure \ref{fig:2} the number of degrees of freedom
for each three-dimensional harmonic oscillator is plotted. As this
figure shows the number of degrees of freedom will be reduced from 6
to 4 in Snyder space. This may be compared with the statistical
mechanics of the ideal gases considered in Ref. \cite{Snyder-IG} in
which the same result is obtained. However, we would like to
emphasize that the reduction of the number of degrees of freedom for
the case of ideal gas can be interpreted as an effective dimensional
reduction of the space at the high temperature regime which is a
common feature of quantum gravity proposals \cite{DR-QG} and also
phenomenological approaches to the minimal length conjecture
\cite{DR-ML}.

Considering the limit $\Theta=\frac{\kappa}{\sqrt{ 2mT}}\gg1$ is
useful at least for two reasons. First, we see that the modified
thermodynamical relations recover the standard results at the
sufficiently low temperature regime when the minimal length effect
are negligible (correspondence principle). Second, one can estimate
the magnitude of the order of the quantum gravity corrections to the
thermodynamical quantities which may potentially lead to observable
effects. In this limit, the series in the relation
(\ref{HO-pf-asym}) converges and one can use it to obtain all the
minimal length (quantum gravity) corrections to the internal energy
and specific heat. Substituting the relation (\ref{HO-pf-asym}) into
(\ref{HO-TPF}) and then using the definition $U=
T^2\left(\frac{\partial\ln{\mathcal Z}}{\partial{T}} \right)_N$, it
is easy to show that the internal energy in the limit of
$\Theta\gg1$ will be
\begin{eqnarray}\label{HO-IE-asym}
\frac{U}{U_0}=1+\frac{2}{3}\sum_{n=1}^{\infty}
(-1)^n\frac{n(2n+1)!!}{\left(2\Theta^2\right)^n},
\end{eqnarray}
where $U_0=3NT$ is the internal energy for the standard nondeformed
$N$ independent three-dimensional harmonic oscillators. In this
limit, the specific heat can be also obtained by substituting the
above relation into the definition $C_{_V}=\left(\frac{\partial{U}}{
\partial{T}}\right)_V$ which gives
\begin{eqnarray}\label{HO-SH-asym}
\frac{C_{_V}}{C_{0_V}}=1+\frac{2}{3}\sum_{n=1
}^{\infty}(-1)^n\frac{n(n+1)(2n+1)!!}{\left(
2\Theta^2\right)^n},
\end{eqnarray}
where $C_{0_V}=\left(\frac{\partial{U_0}}{
\partial{T}}\right)_V=3N$ is the specific heat
of the corresponding nondeformed case.

As is clear from the relations (\ref{HO-IE-asym}) and
(\ref{HO-SH-asym}), the first minimal length (quantum gravity)
corrections to the internal energy and specific heat of the harmonic
oscillator are of the order of $\Theta^{-2}$. To estimate the
magnitude of the order of these corrections, consider the
vibrational oscillations of a carbon monoxide molecule which may be
modeled with the oscillators of mass $m\approx10^{-26}\,kg$ and
frequency $\sigma\approx10^{15}\,Hz$. With this numerical value for
the mass and also considering $\kappa={\mathcal O}(\,1)T_{_{\rm
Pl}}\sim10^{19} \,GeV$ and $T\sim\,1\,TeV$, we get $\Theta\sim
10^{17}$. Therefore, the first minimal length corrections to the
internal energy and specific heat of the harmonic oscillator in
Snyder space are of the order of $\Theta^{-2} \sim10^{-34}$ which is
too small to be experimentally detected for the accessible energy
scales.
\section{Summary}
Appearance of the Hamiltonian systems with nontrivial structure in
the context of phenomenological quantum gravity candidates, such as
the doubly special relativity theories with deformed noncommutative
phase spaces, naively suggests the revision of the statistical
mechanics formalism. In the first step, we have formulated the
statistical mechanics of a general Hamiltonian system in a covariant
(chart-independent) manner by means of the symplectic geometry. The
results show that the two properties of the phase space, as a
symplectic manifold, play distinguished roles in statistical
consideration of a system: The topology of the phase space as a
global property and the local form of the symplectic structure that
determines the geometry of the phase space. For a topologically
trivial phase space with the canonical symplectic structure, the
standard statistical mechanics emerges as it is expected. The
subtleties arise when the phase space has nontrivial topology or
geometry such as the phase spaces with curved momentum space and
noncommutative structure. The topology of the phase space turns out
to be related to the total number of microstates, for instance, the
number of accessible microstates will be finite for a system with
compact phase space. The symplectic structure also affects the
probability distribution of the microstates. The canonical form of
the symplectic structure in terms of the positions and momenta of
the particles plays an important role in the standard statistical
mechanics. It induces a uniform probability distribution over the
microstates by defining a uniform measure on the corresponding phase
space. However, the noncommutative phase space always provides a
noncanonical symplectic structure in terms of the positions and
momenta of the particles and the probability distribution then will
be nonuniform. This is a general result for the phase spaces with
noncanonical (noncommutative) structure which can be immediately
realized from the covariant formulation of the phase space. We
implemented the covariant formalism in order to study the
statistical mechanics in Snyder noncommutative space as an explicit
example of a phase space with nontrivial topology and geometry. We
obtained the associated partition function from which all the
thermodynamical properties of the system can be obtained. As a
particular example, we obtained the partition function for the
three-dimensional harmonic oscillator and we have shown that our
result is in good agreement with that which arises from the full
quantum consideration at the high temperature regime. While there is
no analytical expression for the full quantum partition function of
the harmonic oscillator in Snyder space, the semiclassical
approximation provides an analytical partition function which is
applicable at the high temperature regime. Using the obtained
partition function, we have studied the thermodynamical properties
of the system of harmonic oscillators in Snyder space and our
analysis shows that the number of microstates will be drastically
reduced in this setup due to the existence of a minimal length. This
result justifies our general claim for the deformed spaces which
take into account a minimal length scale. Apart from the details of
the models which deal with a universal minimal length such as the
noncommutative spaces, doubly special relativity theories, the
generalized uncertainty principle, and polymer quantization, the
covariant formalism reveals the main role of the minimal length in
statistical mechanics: The microstates with higher energy or momenta
are less probable when there is a minimal length scale for the
system. Also, we calculated the quantum gravity corrections to the
internal energy and specific heat of a system of harmonic
oscillators. We estimated the order of magnitude of the first
minimal length corrections which are of the order of $10^{-34}$ for
both of the internal energy and specific heat. The equilibrium as an
essential criterion for the statistical system is also considered in
the Appendix by means of the dynamically invariant Liouville volume
which enters in the definition of the Gibbs entropy and partition
function of the statistical system.

\renewcommand{\theequation}{A-\arabic{equation}}
\setcounter{equation}{0}
\section*{APPENDIX: Equilibrium}
\subsection{Statistical independence}
Relation (\ref{decomposition 2}) can be seen as a particular case in
which the system is divided into statically independent subsystems
and subsystems were taken to be the particles themselves. It is
clear that the statistical independence between particles will be
spoiled for a system in which the particles interact with each
other. However, when the system is in equilibrium, one can still
divide it into $M$ macroscopic subsystems ($1\ll M\ll N$) which are
statistically independent. In this respect, the total phase space
$\Gamma$ and the associated symplectic structure $\omega$ will be
\begin{eqnarray}\label{gamma split m}
\Gamma=\Gamma_1\times...\times\Gamma_M\,,\,\,\,\,\,\,\omega=\sum_{A=1}^M
{\omega}_{_A}\,.
\end{eqnarray}
The corresponding Liouville volume is then given by
\begin{align}\label{decomposition 7}
\omega^{3N}=({\omega}_{1})^{3n_1}{\wedge\,.\,.\,.\,\wedge}({\omega}_{M})^{3n_M},
\end{align}
where $n_{_A}$ is the number of particles of the subsystems and is
assumed to be sufficiently large $(n_{_A}\gg 1)$ to guarantee that
the subsystems are macroscopic. The statistical independence of the
subsystems is then defined as
\begin{eqnarray}\label{decomposition 5}
\rho\,\omega^{3N}=\rho^{1}\,({\omega}_{1})^{3n_1}\wedge.\,.\,.\wedge \rho^{M}\,
({\omega}_{M})^{3n_M}\,.
\end{eqnarray}
This criterion is usually employed as an intrinsic characterization of an
equilibrium probability distribution \cite{equilibrium}. Indeed, the system
is in equilibrium if components of the system are in relative equilibrium with
respect to each other. Based on this internal definition of the equilibrium,
the Gibbs state can be understood as an equilibrium state as follows. Consider
the total Hamiltonian function of the system as
\begin{eqnarray}\label{decomposition 8}
H=\sum_{A=1}^M H_A,
\end{eqnarray}
where $\{H_A\}$ are the Hamiltonian functions of the $M$ macroscopic
subsystems and we also disregard the interaction between subsystems
based on the fact that the relative number of particles which take
part in the interactions is negligible compared to $n_{_A}$. So,
using the relations (\ref{can rho class}), (\ref{gamma split m}),
and (\ref{decomposition 8}), the Gibbs state turns out to be an
equilibrium state based on (\ref{decomposition 5}). Additionally, it
is straightforward to show that
\begin{equation}\label{equilibrium}
[{\mathbf x}^{A}_{_H},{\mathbf x}^{B}_{_H}]=0\,,
\end{equation}
which is the criterion for the kinematical and dynamical
separability of the subsystems. The relation (\ref{equilibrium})
then can be viewed as the equilibrium criterion when the system is
in a Gibbs state.

Note that the decomposition (\ref{decomposition 8}) holds only over
not-too-long intervals of time since the effects of interaction of
the subsystems will eventually be dominated even if such
interactions are too weak such as the universal gravitational
effects \cite{padmanabhan}.
\subsection{Liouville equation}
Another more common definition for the equilibrium statistical state of the
system is
\begin{eqnarray}\label{equili}
\frac{\partial\rho_e}{\partial t}=0\,,
\end{eqnarray}
in which $\rho_e$ denotes the density that corresponds to the
equilibrium state. The fact that the Gibbs state is an equilibrium
state then will be justified through the Liouville equation
\begin{eqnarray}\label{Liouville E}
\frac{d}{dt}(\rho\,\omega^{3N})=\Big(\frac{\partial}{\partial t}+
{\mathcal{L}}_{{\mathbf x}_{_H}}\Big)(\rho\, \omega^{3N})=0\,,
\end{eqnarray}
in which the Liouville theorem (\ref{Liouvillee}) is used. The
Liouville equation should be satisfied by any statistical state as
well as Gibbs one and it is, indeed, the necessary requirement for
the consistent probabilistic interpretation of $\rho$. Using the
fact that ${\mathcal{L}}_{{\mathbf x}_{_H}} \rho={{\mathbf
x}_{_H}}\rho=\{\rho,H\}$, the Liouville equation (\ref{Liouville E})
can be rewritten in the more well-known form
\begin{eqnarray}\label{Liouville E2}
\frac{\partial\rho}{\partial t}+\{\rho,H\}=0\,.
\end{eqnarray}
From the relation (\ref{can rho class}), it is clear that
$\{\rho_c,H\}=0$, and the Gibbs state then satisfies the relation
${\partial \rho_c}/{\partial t}=0 $, which shows that it is indeed
an equilibrium state through the definition (\ref{equili}).

If we use the Liouville equation (\ref{Liouville E}) and consider
the fact that the Liouville volume $\omega^{3N}$ does not change
explicitly with time, it turns out that
\begin{eqnarray}\label{dt entropy}
\frac{dS}{dt}=-\frac{1}{N!}\Big(\frac{\partial}{\partial t}+
{\mathcal{L}}_{{\mathbf x}_{_H}}\Big)\int_{\Gamma}\rho \ln{\rho}
\,\omega^{3N}\nonumber\\=\frac{1}{N!}\int_{\Gamma}\rho\,
{\mathcal{L}}_{{\mathbf x}_{_H}}\omega^{3N}=0\,,\hspace{1.3cm}
\end{eqnarray}
which shows that the fine-grained entropy (\ref{gibbs entropy}) does
not change with time as expected for closed systems. The above
relation and the Liouville equation (\ref{Liouville E}) guarantee
that the number of microstates does not change through the dynamical
evolution of the system, which makes the statistical formulation a
consistent setup to give the thermodynamical properties in any
ensemble.


\begin{thebibliography}{1}
\bibitem{unit} We work in units $\hbar
=c=k_{_B}=1$, where $\hbar$, $c$ ,
and $k_{_B}$ are the Planck constant,
speed of light in vacuum, and
Boltzmann constant respectively.
\end{thebibliography}
\end{document}